\documentclass{article}

\PassOptionsToPackage{compress,sort}{natbib}

\usepackage[utf8]{inputenc} 
\usepackage[T1]{fontenc}    
\usepackage[margin=1in]{geometry}
\usepackage{hyperref}       
\usepackage{url}            
\usepackage{booktabs}       
\usepackage{amsfonts}       
\usepackage[natbibapa]{apacite}
\usepackage{nicefrac}       
\usepackage{microtype}      
\usepackage{xcolor}         

\usepackage{amsmath,amssymb,amstext}
\usepackage{graphicx}
\usepackage{mathrsfs}
\usepackage{subcaption}

\newcommand{\argmax}{\mathop{\mathrm{argmax}}}
\newcommand{\cond}[2]{#1 \vert #2}

\newcommand{\Misspecification}{\mathscr{D}_{model}}
\newcommand{\Expect}[2]{\mathbb{E}_{#1}\left[ #2 \right]}
\newcommand{\SampledX}{\mathbf{x}}
\newcommand{\SampledY}{\mathbf{y}}
\DeclareMathOperator*{\argmin}{argmin} 
\newcommand{\KLD}{$D_{KL}$}
\newcommand{\KLDfromto}[2]{D_{KL} \left( #1 ~ \vert \vert ~ #2 \right)}

\newcommand{\TargetDist}{g}
\newcommand{\TrueConditional}{f}
\newcommand{\Posterior}{p}
\newcommand{\Model}{m}
\newcommand{\SampledXBeforeT}{\SampledX_t}
\newcommand{\SampledYBeforeT}{\SampledY_t}
\newcommand{\Utility}{\Expect{}{u_{t+1}(x)}}

\definecolor{cb1}{RGB}{27,158,119}
\definecolor{cb2}{RGB}{217,95,2}
\definecolor{cb3}{RGB}{117,112,179}

\title{Characterizing the robustness of Bayesian adaptive experimental designs to active learning bias}

\begin{document}

\begin{center}
    \LARGE{
        Characterizing the robustness of Bayesian adaptive experimental designs to active learning bias
    }
    
    \vspace{8mm}
    
    \normalsize{

        \begin{tabular}{ccc}
            Sabina J.~Sloman & & Daniel M.~Oppenheimer \\
            Department of Social and Decision Sciences & & Departments of Social and Decision Sciences and of Psychology \\
            Carnegie Mellon University & & Carnegie Mellon University \\
            Pittsburgh, PA 15213 & & Pittsburgh, PA 15213 \\
            \texttt{ssloman@andrew.cmu.edu} & & \texttt{oppenheimer@cmu.edu} \\
            & & \\
            & & \\
            Stephen B.~Broomell & & Cosma Rohilla Shalizi \\
            Department of Psychological Sciences & & Departments of Statistics and of Machine Learning \\
            Purdue University & & Carnegie Mellon University \\
            West Lafayette, IN 47907 & & Pittsburgh, PA 15213 \\
            \texttt{broomell@gmail.com} & & Santa Fe Institute \\
            & & Santa Fe, NM 87501 \\
            & & \texttt{cshalizi@cmu.edu}
        \end{tabular}

        \vspace{8mm}

        \today{}
    }

\end{center}

\vspace{8mm}

\begin{abstract}
  Bayesian adaptive experimental design is a form of active learning, which chooses samples to maximize the information they give about uncertain parameters.  
  Prior work has shown that other forms of active learning can suffer from {\bf active learning bias}, where unrepresentative sampling leads to inconsistent parameter estimates.  
  We show that active learning bias can also afflict Bayesian adaptive experimental design, depending on model misspecification. 
  We analyze the case of estimating a linear model, and show that worse misspecification implies more severe active learning bias.  
  At the same time, model classes incorporating more ``noise'' --- i.e., specifying higher inherent variance in observations --- suffer less from active learning bias.
  Finally, we demonstrate empirically that insights from the linear model can predict the presence and degree of active learning bias in nonlinear contexts, namely in a (simulated) preference learning experiment.
\end{abstract}

\section{Active learning}\label{sec:al}
    Statistical theory often assumes learners' access to large amounts of representative training data, drawn from the distribution which is the target of inference or prediction.
    Nonetheless, such access is not feasible for many applications.
    Training data may be scarce (e.g., learning to identify a rare medical condition; \citet{henry_targeted_2015}), difficult or expensive to obtain (e.g., requiring human coders for text; \citet{chen_study_2015}), or time-consuming to collect (e.g., obtaining user preferences online; \citet{golovin_near-optimal_2010,cavagnaro_optimal_2013}).
    One response is to abandon random sampling for adaptive sampling methods, choosing data points in sequence to be as informative as possible.
    In many settings, such \textbf{active learning} or \textbf{adaptive sampling} methods let us make strong inferences or achieve high rates of out-of-sample accuracy while saving enormously on training data.\footnote{
        We will use ``adaptive sampling'' and ``active learning'' interchangeably.
    }

    We focus on a class of active learning methods called \textbf{information theoretic active learning}, where the modeler must pick an objective function capturing the informativeness of observations \citep{houlsby_collaborative_2011}.
    The amount of information contained in any given observation can be quantified using information theory \citep{lindley_measure_1956,cover_elements_1991}.
    The general framework is highly extensible, and objective functions can encompass almost anything one might want to maximize information \emph{about} \citep{lindley_measure_1956,chaloner_bayesian_1995,ryan_review_2016}.
    In active learning, one usually seeks observations that will maximally reduce uncertainty about parameters \citep{houlsby_collaborative_2011}.

    Beyond machine learning, these methods are also widely used in the context of experimental design, in fields ranging from neuroscience \citep{lewi_sequential_2009}, cognitive psychology \citep{cavagnaro_adaptive_2010}, and clinical medicine \citep{giovagnoli_bayesian_2021} to computational physics \citep{huan_simulation-based_2013}.
    Information theoretic active learning methods are used to identify powerful experimental designs, which can facilitate rapid scientific understanding by efficiently estimating interpretable parameter values or recovering core mechanistic structure of a system.

    We thus refer to the particular class of information-theoretic active learning algorithms studied in this paper, described in detail in \S \ref{sec:baedo}, as \textbf{Bayesian adaptive experimental design}, or simply \textbf{Bayesian adaptive design}.

    \paragraph{Active learning bias}
        Despite their advantages, adaptive sampling schemes not only introduce sequential dependencies between observations (breaking IID assumptions), but can also produce training sets that are highly unrepresentative of the target distribution \citep{farquhar_statistical_2021}.
        This implies that the estimates made from actively-sampled data may not generalize to the target.
        This phenomenon is {\bf active learning bias} \citep{farquhar_statistical_2021}.
        While theoretically concerning, in practice active learning bias does not \emph{always} happen \citep{varghese_active_2019, farquhar_statistical_2021}.
        Understanding when and how active learning bias leads to poor generalization is thus an important area of inquiry within adaptive sampling.

    \paragraph{Model misspecification}\label{sec:model-misspecification}
        The advantages of active learning methods are usually expounded assuming that the model, or hypothesis, class is well-specified, i.e., that the true data-generating distribution is a member of the class \citep{mackay_information-based_1992,kanamori_statistical_2002,dasgupta_analysis_2004,sugiyama_active_2005,myung_tutorial_2013}.
        Yet in most modeling enterprises, this assumption is not credible: the exact form of the true data-generating process is difficult if not impossible to know, and models are deliberately simplified tractable approximations.
        This is of especial concern in the experimental paradigms mentioned above, in which the aim is usually to \emph{learn} an interpretable approximation to the generative process.
        Even in the best case, core aspects of the generative process are unknown to or ignored by the model (while in the worst case, the model, or current state of scientific knowledge, may be more simply mistaken about the generative process).

        While sampling bias and vulnerability to model misspecification are often discussed as two separate limitations of active learning methods, it turns out they are deeply related.
        In particular, when the model class is misspecified, sampling bias, a form of covariate shift, can amplify bias \citep{sugiyama_direct_2008,wen_robust_2014,spencer_feedback_2021}.

\begin{figure}[t!]
            \centering
	        \includegraphics[width=.75\linewidth]{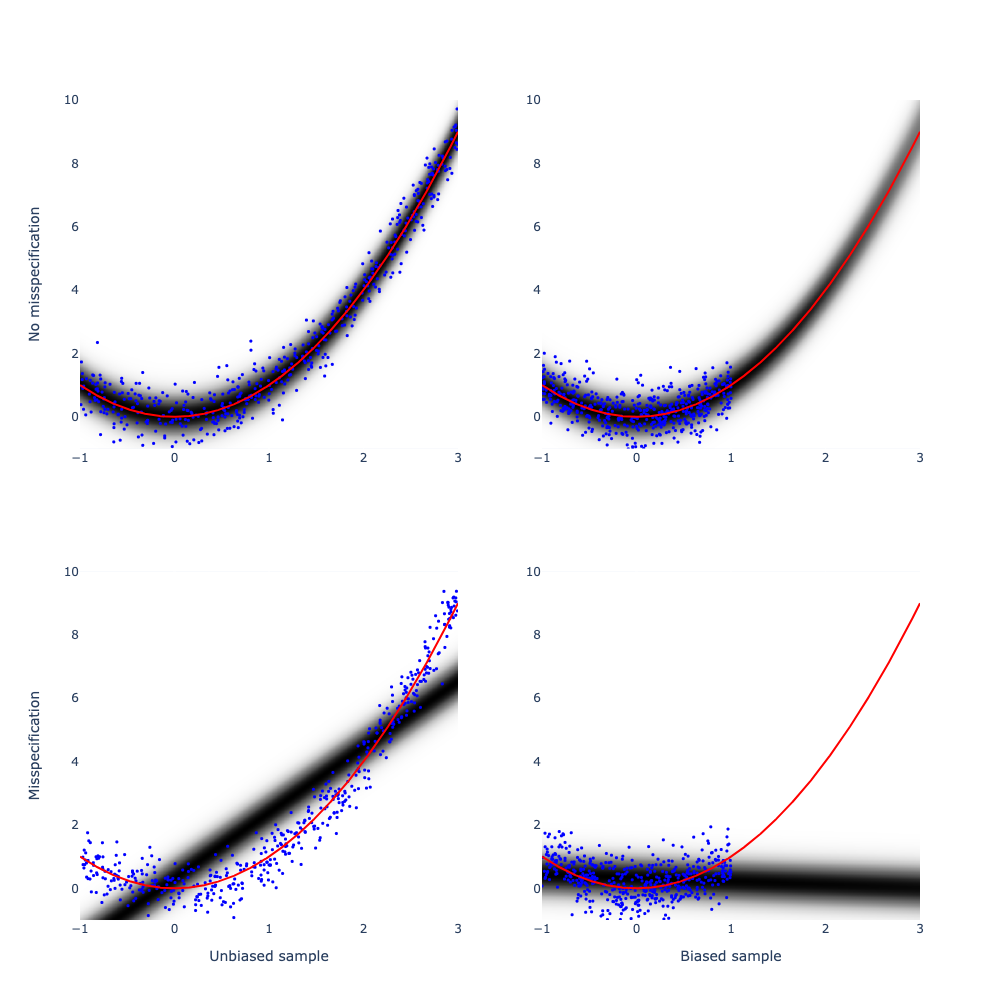}
		    \caption[The effect of model misspecification and sampling bias: Motivating example.]{The effect of model misspecification and sampling bias: Motivating example (see discussion in the main text).}
		    \label{fig:misspecification}
	    \end{figure}

        To build intuition, look at Figure \ref{fig:misspecification}.
        In all four panels, the true model generating the data, whose expectation is shown by the red lines, is $y|x \sim \mathscr{N}(x^2, .5)$.
        In the top panels, the class of models, or hypotheses, considered is $y|x \sim \mathscr{N}(\beta x^2, .5)$, so the model class is well-specified.
        In the bottom panels, the class of models considered is $y|x \sim \mathscr{N}(\beta x, .5)$, which is misspecified.

        The blue dots show the training data points.
        If the target distribution is uniformly distributed along the $x$ axis, the samples shown in the lefthand panel are representative --- they are drawn from the target distribution --- and the samples shown in the righthand panel are not --- they are restricted to only part of the domain of the target distribution.

        The predictive distributions obtained from training on each data set are shown in the corresponding panel by gray shading; darker areas indicate higher predicted probability.
        In the top panels, where the model class is well-specified, the introduction of sampling bias does not impair the model's ability to capture unseen data.
        However, in the bottom panels, where the model class is misspecified, introducing sampling bias \emph{does} degrade generalization: the model in the bottom right panel generalizes much more poorly across the target distribution \emph{because of} the introduced sampling bias.
        Intuitively, the well-specified model in the top right panel is able to use its structure to constrain itself appropriately even in areas of the target distribution it hasn't seen \citep{sloman_you_2020}.
        On the other hand, the misspecified model in the bottom right panel has constrained itself inappropriately.

        The implication of this is that when the model class is misspecified, active learning bias can result \citep{sugiyama_active_2005,bach_active_2006}.

\section{Goals of the present work}
    The goals of this paper are 1) to apply insights on active learning bias from other active learning paradigms to Bayesian adaptive design, and 2) to investigate the conditions under which model misspecification leads to active learning bias.
    Despite the formal connections between active learning and adaptive experimental design, to the best of our knowledge, the phenomenon of active learning bias in the context of Bayesian experimental design has yet to be studied explicitly.
    Such a study has practical import, especially given the use of Bayesian adaptive design as a tool for individualized assessment and scientific inference.

    In addition, our work extends inquiry of the conditions under which active learning bias occurs, yielding results on the mediating role of \emph{degree of model misspecification}, and practical recommendations of how to mitigate active learning bias.
    In particular, our work:

    \begin{enumerate}
        \item demonstrates that model misspecification predicts the presence of active learning bias,
        \item investigates how the extent of this bias varies with the degree of model misspecification, and
        \item shows how model specifications that predict more inherent observation variance mitigate active learning bias.
    \end{enumerate}

\section{Preliminaries and notation}
    A modeler wants to predict some variable $y \in \mathbb{R}^{m}$ using another variable $x \in \mathbb{R}^d$.
    Given $x$, $y$ always follows the same distribution, $y|x \sim \TrueConditional(x)$.
    We call $\TrueConditional$, the generating conditional distribution, the {\bf true model}.
    We assume that $x$, the input or \textbf{design}, is fully observable, always available before $y$ must be predicted, and that it follows some distribution $\TargetDist$.

    The modeler specifies a {\bf hypothesized model class} that predicts $y|x \sim \Model(x, \theta)$, with $\theta \in \Theta$; the variable $\theta$ contains the {\bf parameters} of the model class, which live in the parameter space $\Theta$.
    $\Theta$ is fixed, i.e. determined in advance of the data and unchanging in response to them.
    $\Model$ is a probabilistic function, whose form is also presumed to be fixed (e.g., logistic regression).
    The hypothesized model class $\Model(x, \Theta)$ is thus comprised of the set of distributions $\{ \Model(x, \theta) ~ : ~ \theta \in \Theta \}$.

    The model estimation problem is to find $\theta^* \in \Theta$ which minimizes a loss function that captures the modeler's objective, i.e., the risk.
    In other words, the risk minimizer $\theta^*$ is defined as:

    \begin{equation}\label{eq:thetastar}
        \theta^* \equiv \argmin_{\theta \in \Theta}{ \Expect{}{\mathscr{L}(\Model(x, \theta), y)}}
    \end{equation}

    for some suitable loss function $\mathscr{L}$, which takes as inputs a predictive distribution and a realized value for $y$.
    Here, the expectation is under the true data-generating distribution, i.e., the product of $\TargetDist$ and $\TrueConditional$.
    The function $\Model(x, \theta^*)$ is an instance of the hypothesized model class, the {\bf best-fitting model}.\footnote{
        Some related work writes of selecting a \emph{hypothesis} from a \emph{hypothesis class} \citep{dasgupta_analysis_2004,golovin_near-optimal_2010}.
        Our ``hypothesized model class'' matches their ``hypothesis class,'' and selecting $\theta^*$ from $\Theta$ amounts to selecting a hypothesis from the hypothesis class.
        Our terminology is similar to \cite{atmanspacher_model_2016}, who also distinguish between model classes and model instances in the context of Bayesian modeling.
    }

    To estimate $\theta^*$, the modeler gets a set of samples $\SampledX \in \mathbb{R}^{n \times d}$ and observes outcomes $\mathbf{y} \in \mathbb{R}^{n\times m}$.
    We consider only distributional estimates of $\theta^*$, essentially assigning a probability to each $\theta \in \Theta$ that it is the risk minimizer:

    \begin{equation}
        \Posterior(\theta) \equiv P(\theta = \theta^*|\mathbf{x}, \mathbf{y}).
    \end{equation}

    The corresponding predictive distribution for $y|x$ is

    \begin{equation}
        y|x \sim \int_{\theta}{\Model(x, \theta) ~ \Posterior(\theta) ~ d\theta}.
    \end{equation}

    For brevity, we call this $\Model(x, \hat{\theta})$, though there need not be a single $\theta \in \Theta$ giving exactly this distribution.
    We call $\Model(x, \hat{\theta})$ the {\bf trained} or {\bf estimated} model.

    Efficient model estimation concentrates $\Posterior$ on a small set with relatively few observations (i.e., with small $n$).
    Random sampling or passive learning techniques draw IID $x$ values from the population distribution $\TargetDist$.
    In other words, the sampling or \textbf{design distribution} is unbiased; asymptotically, it will converge to $\TargetDist$.
    Adaptive sampling or active learning techniques \emph{actively} construct $\SampledX$ to maximize the concentration of $\Posterior$.
    Usually, this results in sampling bias; the design distribution diverges from $\TargetDist$.

    Both random and adaptive sampling then result in observations $\SampledY|\SampledX \sim \TrueConditional(\SampledX)$.
    We write $\Model(x, \hat{\theta}_{adaptive})$ for the result of an adaptive procedure, and $\Model(x, \hat{\theta})$, unmodified, for passive learning.

    Note that risk, the modeler's objective, is defined using the distribution $\TargetDist$, the {\bf target distribution} of inputs whose consequences the modeler ultimately wants to predict.
    Active learning bias (ALB) occurs when, averaging over data sets $x \sim \TargetDist$ and corresponding observations $y|x\sim \TrueConditional$, $\Expect{}{\mathscr{L}(\Model(x, \hat{\theta}_{adaptive}), y)} > \Expect{}{\mathscr{L}(\Model(x, \theta^*), y)}$.
    In general, this bias can be sample-size dependent; our goal is to approximate its properties as $n\rightarrow\infty$.

    We say that a hypothesized model class is {\bf misspecified} when the true model is not in the class, i.e., $\TrueConditional(x) \neq \Model(x, \theta)$ for all $\theta \in \Theta$.
    In particular, then, $\Model(x, \theta^*) \neq \TrueConditional(x)$.
    Consistent with prior literature, we say ``model misspecification,'' even though it is the hypothesized model {\em class} that is wrong.

\section{Bayesian adaptive experimental design}\label{sec:baedo}
    A natural way to construct $\Posterior$ is via Bayesian inference.
    We begin with a prior distribution $\Posterior_0$, and, at the $t^{\mathrm{th}}$ step, where we observe $x_t$ and $y_t$, we use Bayes's rule to update this distribution:

    \begin{equation}\label{eq:bayes}
        \Posterior_{t}(\theta) = \Posterior_{t-1}(\theta)\frac{\Model(y_t | x_t, \theta)}{\int_{\theta}{\Model(y_t | x_t, \theta) ~ \Posterior_{t-1}(\theta) ~ d\theta}}.
    \end{equation}

    Here, $\Model(y_t | x_t, \theta)$ indicates the likelihood of observation $y_t$ under the distribution $\Model(x_t, \theta)$.
    $\Posterior_{t}(\theta)$ thus implicitly involves the whole history of inputs and responses to date, $\SampledXBeforeT$ and $\SampledYBeforeT$ respectively.

    For Bayesian adaptive experimental design, at every step $t$ we define {\em the modeler's} expected utility on the following trial of an input $x$ over the current posterior distribution of $\theta$:

	  \begin{equation}\label{eq:Eu}
		  \Expect{}{u_{t+1}(x)} = \int_\theta{ \int_y{ u(x, y, \theta) ~ \Model(y | x, \theta) ~ \Posterior_t(\theta) ~ dy} ~ d\theta}.
	  \end{equation}

    Bayesian adaptive experimental design picks the maximizer of $\Expect{}{u_{t+1}(x)}$ as the next value of $x$, or the \textbf{optimal design}.

    The utility function $u$ encodes the goals of the experiment.
    We pick a $u$ that, motivated by information-theoretic considerations, encourages precision of parameter estimates:

    \begin{equation}\label{eq:u}
	    u(x, y, \theta) = \log{\frac{\Posterior_t(\theta | y, x)}{\Posterior_t(\theta)}}.
    \end{equation}

    (To be clear, the numerator in the ratio is the posterior probability that $\theta$ {\em would} have, if we should happen to observe $x$ and $y$.)

    With this choice of $u$, $\Expect{}{u_{t+1}(x)}$ is the mutual information between the next observation and the parameter $\theta$, regarded as a random variable distributed according to the posterior distribution \citep{bernardo_expected_1979}.
    This criterion is commonly referred to as the \emph{expected information gain (EIG)}, and is the dominant criterion for Bayesian adaptive design in many fields, particularly in the behavioral sciences \citep{myung_tutorial_2013,foster_variational_2021}.
    Note that EIG is also equivalent to other commonly-used objective functions for Bayesian active learning, e.g., expected decrease in posterior entropy \citep{houlsby_collaborative_2011} and expected Kullback-Leibler divergence between prior and posterior parameter estimates \citep{huan_sequential_2016}.

\section{Measuring model misspecification}\label{sec:measure-misspecification}
    We want to investigate how active learning bias varies with the degree of misspecification.
    We hypothesize that the degree of misspecification will correlate positively with the degree of active learning bias.
    The intuition for this comes from the dependence of the design distribution on $\Posterior(\theta)$.
    We intuit that the more mistaken the estimation process is about the generating model, the more mistaken it will be about where in the design space is most informative to query.

    Luckily, there is a natural information-theoretic way to measure degree of misspecification, using the Kullback-Leibler divergence (\KLD{}).
    Concretely, we define the degree of misspecification as the expectation, under $\TargetDist$, of the \KLD{} between the true $\TrueConditional$ and the best-fitting model:

    \begin{equation}\label{eq:misspecification}
        \Misspecification \equiv \int_x{ \KLDfromto{\TrueConditional(x)}{\Model(x, \theta^*)} ~ g(x) ~ dx}.
    \end{equation}

    Standard results in Bayesian theory \citep{berk_limiting_1966,shalizi_dynamics_2009} say that the \KLD{} between $\TrueConditional$ and the modeler's posterior predictive distribution will approach $\KLDfromto{\TrueConditional(x)}{\Model(x, \theta^*)}$ as $n\rightarrow\infty$ under passive learning,\footnote{
        So long as $\mathscr{L}$, the loss function whose expectation $\theta^*$ minimizes (see Equation \ref{eq:thetastar}), is a proper scoring rule.
    } so this is a reasonable measure of the systematic error inherent in hypothesizing a particular model class.

\section{Bayesian adaptive design for linear models}\label{sec:linear-ex}
    Our first set of results focuses on the case of Bayesian linear regression with fixed observation variance.
    We first explain the setup in detail, and discuss two important properties of Bayesian adaptive design for such models: The optimal design is independent of the history of observations, but depends on the assumed observation variance.

    Prior work has demonstrated that in the linear regression case, batch optimal designs \emph{do} induce bias in the presence of model misspecification \citep{box_basis_1959,myers_response_1989,shi_model-robust_2021}.
    As discussed in detail in the following section, in this paradigm, the batch optimal design is the same as the optimal adaptive design.
    Thus, these prior results hold for the adaptive case as well.
    The empirical results presented in \S \ref{sec:alb-exists} verify that ALB occurs in the presence of model misspecification.
    \S \ref{sec:degree-misspecification} and \S \ref{sec:noise} present two novel findings:

    \begin{itemize}
        \item The degree of ALB depends on the degree of model misspecification.
        \item Increasing the amount of hypothesized observation variance leads to a lower degree of ALB.
    \end{itemize}

    In \S \ref{sec:additional-ex} and Appendix \ref{ap:classification}, we show that all three findings hold in a non-linear environment.

    \subsection{Bayesian linear regression}\label{sec:bayes-linreg}
        We consider a class of polynomial linear regression models where observations $y$ are assumed to be distributed

        \begin{align}
            y \sim \mathscr{N}(\beta\phi^k(x)^T, \sigma^2)
        \end{align}

        where $\phi^k(x)$ indicates a degree-$k$ polynomial expansion of the design $x$.
        The degree $k$ and the observational variance $\sigma^2$ are fixed in advance, while the coefficient vector $\beta \in \mathbb{R}^{k+1}$ is the set of parameters to be learned.

        The vector $\beta$ is assigned a prior distribution

        \begin{align}
            p_0(\beta) &= \mathscr{N}(M_0, S_0)
        \end{align}

        where $M_0$ and $S_0$ respectively denote the prior mean and covariance of the distribution.\footnote{
            Notation borrowed from \citet{bishop_pattern_2006}.
        }

        After an observation $\langle x, y \rangle$,  $M$ and $S$ update as follows:

        \begin{align}\label{eq:linear-sigmat2}
            S_t &= \left( S_{t-1}^{-1} + \frac{\phi^k(x)^T\phi^k(x)}{\sigma^2} \right) ^{-1}
        \end{align}

        and

        \begin{align}
            M_t &= S_t \left( S_{t-1}^{-1}M_{t-1} + \frac{\phi^k(x)^Ty}{\sigma^2} \right).
        \end{align}

        \paragraph{Design criterion}
            The utility function is given by:

            \begin{align}\label{eq:linear-Eu}
                \Expect{}{u_{t+1}} &= \frac{1}{2}\log{\left( \sigma^2 + \phi^k(x)^TS_t\phi^k(x) \right)} - \log{\left( \sigma \right)}.
            \end{align}

            See Appendix \ref{ap:linear-Eu} for the derivation of Equation \ref{eq:linear-Eu}.

            Note that maximizing Equation \ref{eq:linear-Eu} is equivalent to maximizing

            \begin{align}\label{eq:linear-argmaxEu}
                \argmax_x \Expect{}{u_{t+1}} &= \argmax_x \phi^k(x)^TS_t\phi^k(x).
            \end{align}

            Maximizing Equation \ref{eq:linear-argmaxEu} is equivalent to maximizing the Bayesian $D$-optimality criterion \citep{chaloner_bayesian_1995}.

        \paragraph{Independence of the optimal design on the history of observations.}\label{ap:linearu}
            It turns out that the optimal design does not depend on the history of observed $y$ values $\textbf{y}_t$.
            This can be seen by noticing that both $\phi^k(x)$ and $S_t$ are independent of $y_{t-1}$ for all $t$ (see Equation \ref{eq:linear-sigmat2}).
            In other words, the experimenter does not need to actually run the experiment to construct an optimal sequence of samples \citep{mackay_information-based_1992,krause_nonmyopic_2007}.
            In fact, identifying the optimal design after each sequential observation yields (asymptotically) the batch $D$-optimal design \citep{wynn_sequential_1970}.

            The implication of this is that unlike in the non-linear case, the distribution of designs is only a function of the hypothesized model $m$ and prior parameter distribution $p_0$.
            Critically, the distribution of designs will \emph{not} depend on the extent of model misspecification, or indeed, whether the model is misspecified at all.

            On the one hand, this provides us with a ``cleaner'' initial paradigm to study, and simplifies some of our interpretations (e.g., the effect of the extent of model misspecification on ALB; \S \ref{sec:degree-misspecification}).
            On the other hand, it limits the generalizability of the findings from this particular modeling paradigm.
            To address this, in \S \ref{sec:additional-ex} we show that our findings replicate in a non-linear modeling paradigm.

        \paragraph{Dependence of optimal design on $\sigma$.}
            Equation \ref{eq:linear-argmaxEu} shows that the optimal design is the one that maximizes $\phi^k(x)^TS_t\phi^k(x)$.
            This is affected by $\sigma$ through the dependence of $S_t$ on $\sigma$: Equation \ref{eq:linear-sigmat2} shows that the posterior variance of parameter values increases with the assumed variance of observations.

\begin{figure}[h!]
            \begin{subfigure}{.48\linewidth}
                \includegraphics[width=\linewidth]{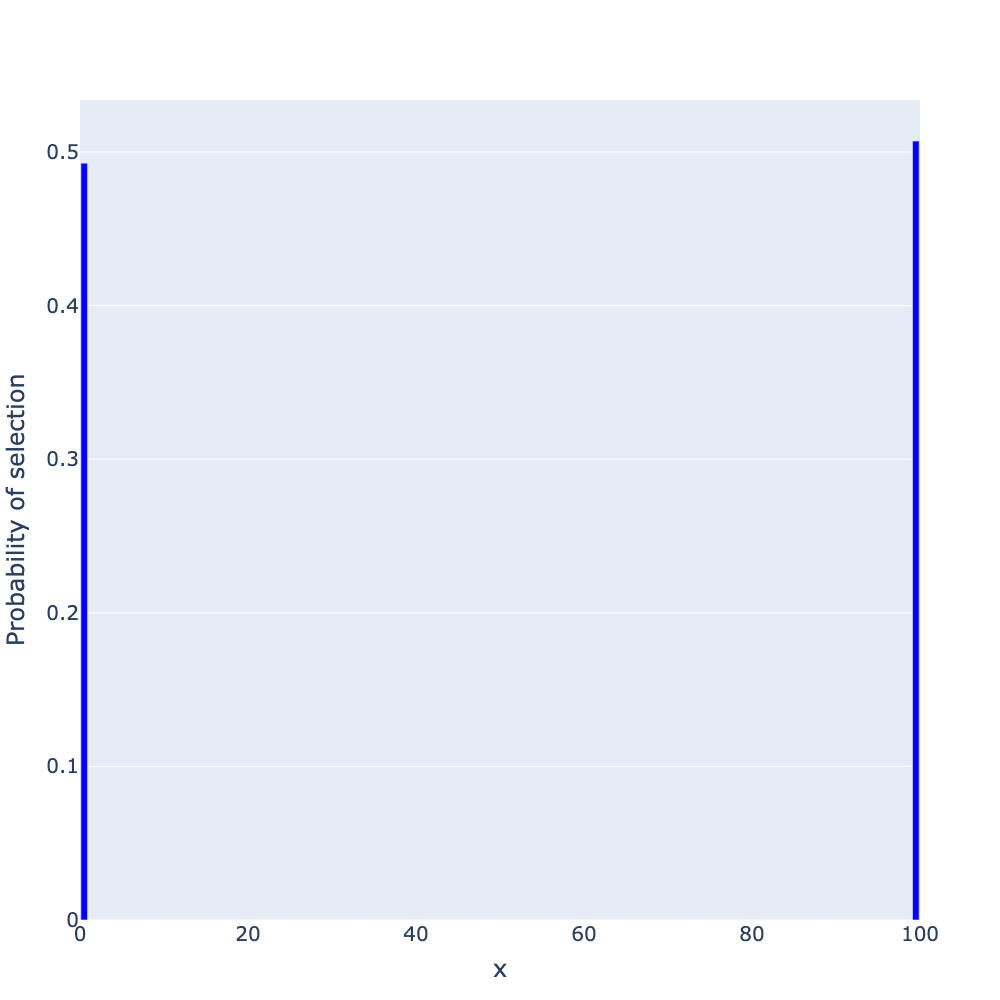}
                \caption{$k = 1, \sigma = 100$.}
                \label{fig:designs-12}
            \end{subfigure}\hfill\begin{subfigure}{.48\linewidth}
                \includegraphics[width=\linewidth]{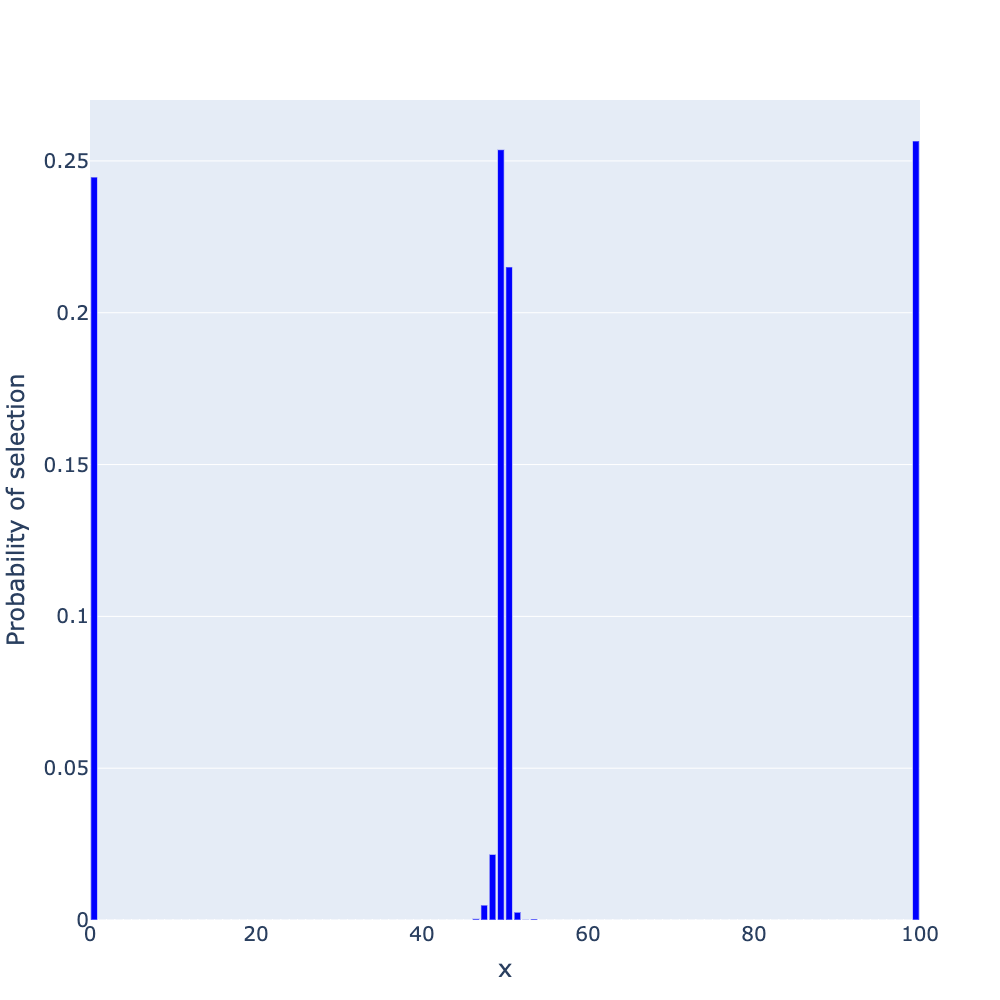}
                \caption{$k = 2, \sigma = 100$.}
                \label{fig:designs-23}
            \end{subfigure}
            \begin{subfigure}{.48\linewidth}
                \includegraphics[width=\linewidth]{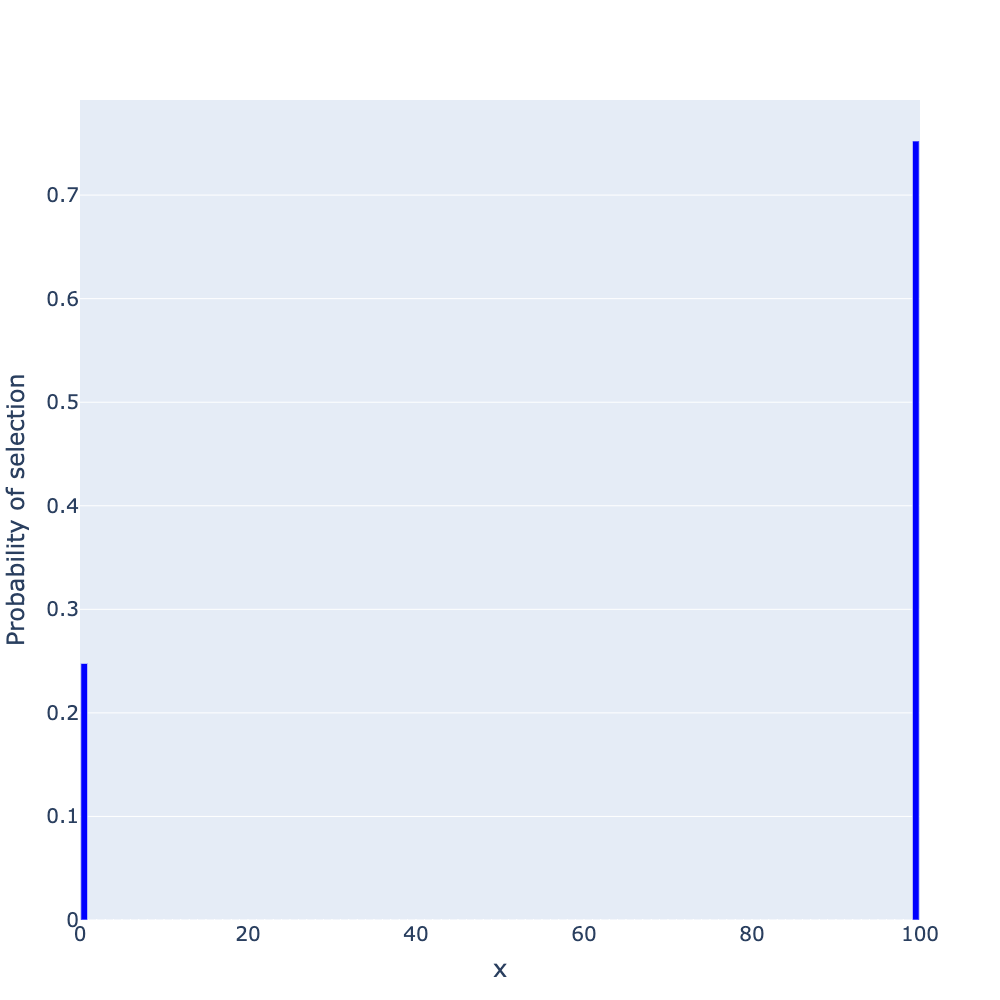}
                \caption{$k = 1, \sigma = 1,000$.}
                \label{fig:designs-12-noise}
            \end{subfigure}\hfill\begin{subfigure}{.48\linewidth}
                \includegraphics[width=\linewidth]{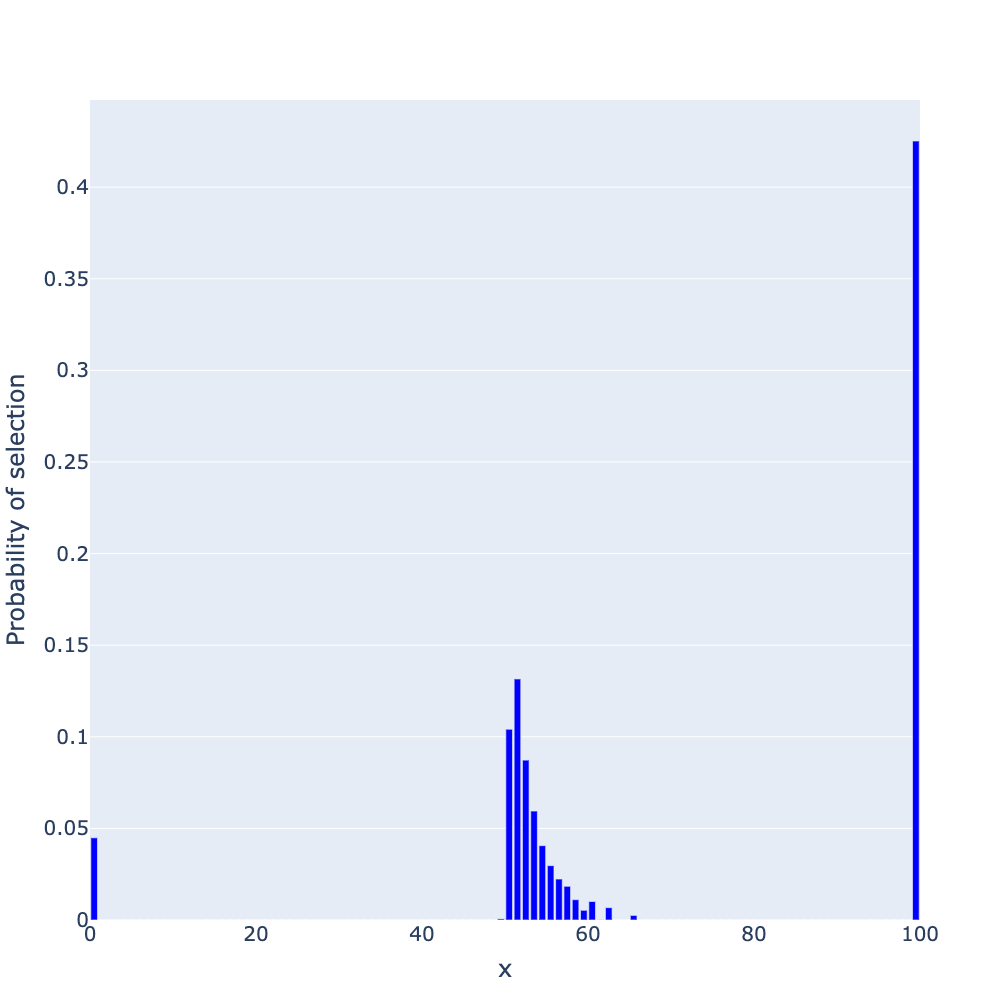}
                \caption{$k = 2, \sigma = 1,000$.}
                \label{fig:designs-23-noise}
            \end{subfigure}
            \caption[Linear regression: Distribution of designs selected by adaptive sampling method.]{Linear regression: Distribution of designs selected by adaptive sampling method.
                Subfigures show distributions characterized by values of $k$ (the degree of the hypothesized polynomial regression model) and $\sigma$ (the degree of hypothesized noise).
                Results from the corresponding simulation experiments are given in sections \ref{sec:alb-exists}--\ref{sec:noise}.
                Each subfigure shows the distribution across designs selected in each of 100 trials $\times$ 1,000 simulation experiments = 100,000 designs.
            }
            \label{fig:designs}
        \end{figure}

        \paragraph{}
            Figure \ref{fig:designs} shows the distribution of designs selected by Bayesian adaptive design across the simulation experiments reported in the following subsections.
            Consistent with the discussion above, the empirical design distributions exhibit dependence on $k$ (the degree of the hypothesized polynomial regression model) and $\sigma$ (the degree of hypothesized noise).

    \subsection{Result 1: ALB depends on model misspecification.}\label{sec:alb-exists}
        This section corroborates results from prior work that bias in $\hat{\beta}$ induced by learning on the optimal sequence of designs depends on model misspecification.

\begin{figure}[h!]
	        \begin{subfigure}{.32\textwidth}
	            \includegraphics[width=\linewidth]{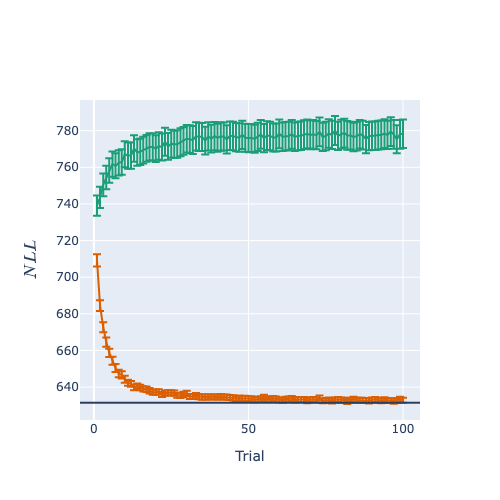}
		        \caption{$\Model$ is linear.}
		        \label{fig:linear-1}
	        \end{subfigure}\begin{subfigure}{.32\textwidth}
	            \includegraphics[width=\linewidth]{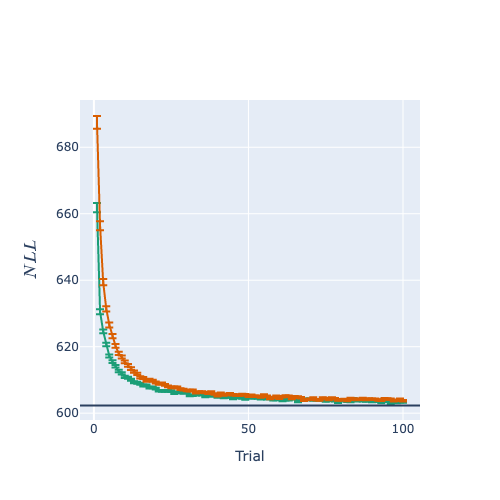}
		        \caption{$\Model$ is quadratic.}
		        \label{fig:linear-2}
	        \end{subfigure}\begin{subfigure}{.32\textwidth}
	            \includegraphics[width=\linewidth]{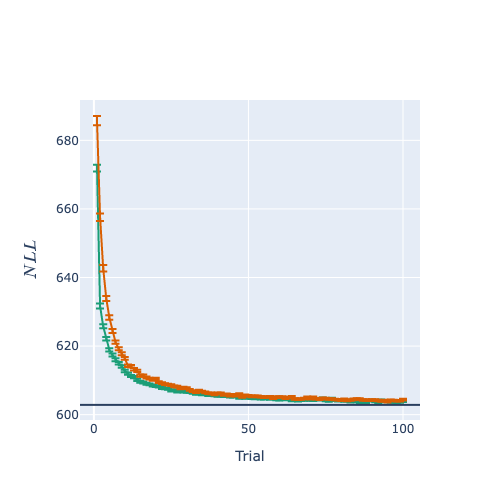}
		        \caption{$\Model$ is cubic.}
		        \label{fig:linear-3}
	        \end{subfigure}
	        \caption[Linear regression: Empirical result 1.]{
            Linear regression: Empirical result 1 (ALB depends on model misspecification).
            Each panel shows the risk incurred across 100 steps of Bayesian updating.
	          Risk is approximated as the negative log likelihood ($NLL$) of 100 observations from the target distribution.
		        The generating model is always a degree-two polynomial with parameters drawn from a $\mathscr{N}([0,0,0], \mathrm{diag}(100, 10, .1))$ distribution.
		        Lines are means across 1,000 simulated experiments, with error bars showing $\pm 1$ standard error around the mean.
		        Horizontal lines show the risk achievable by $\Model(x, \theta^*)$.
		        Green lines ({\color{cb1} \rule{4mm}{1mm}}) show results using Bayesian adaptive design during model estimation.
		        Orange lines ({\color{cb2} \rule{4mm}{1mm}}) show results using random sampling from the target distribution.
	        }
	        \label{fig:linear}
        \end{figure}

        Figure \ref{fig:linear} shows the results of several simulations in which $\TrueConditional$ is a degree-two polynomial regression model, i.e., $y = \beta_0 + \beta_1 x + \beta_2 x^2 + \epsilon$ where $\epsilon \sim \mathscr{N}(0, 100)$.
        In the underparameterized case (leftmost panel of Figure \ref{fig:linear}), $\Model$, the functional form corresponding to the hypothesized model class, is linear in $x$.
        In the fully parameterized case (middle panel), $\Model$ is quadratic in $x$.
        In the overparameterized case (rightmost panel), $\Model$ is cubic.
        In all cases, the additive noise is (correctly) specified as $\epsilon \sim \mathscr{N}(0, 100)$.
        Thus, by our definition, only the underparameterized case is misspecified.

	    Figure \ref{fig:linear} shows that the behavior in the misspecified (underparameterized) case is markedly different from the behavior in the other two cases.
	    Risk was measured as the negative log likelihood ($NLL$) of 100 observations drawn randomly from the target distribution, $\TargetDist = \mathrm{Unif}(0, 100)$.
	    For the well-specified model classes, adaptive sampling reduces the risk more quickly than random sampling, though both converge on the same risk as $\theta^*$.\footnote{
	        We found $\theta^*$ by taking the OLS solution to a regression of 1,001 evenly spaced points across the domain and their expectation under the true model.
	    }
	    In other words, when the model class is well-specified, Bayesian adaptive designs efficiently converge on a model that generalizes across the target distribution.
	    However, under misspecification, Bayesian adaptive designs lead to worse generalization than random sampling.\footnote{
	        Simulations were run with the help of the GNU Parallel program \citep{tange_gnu_2021}.
	    }

    \subsection{Result 2: The \emph{degree of} ALB depends on the \emph{degree of} misspecification.}\label{sec:degree-misspecification}
        We now investigate how the extent of ALB varies with the degree of misspecification (as defined in \S \ref{sec:measure-misspecification}).
        We define ALB as

        \begin{equation}
            ALB \equiv \frac{\Expect{}{\mathscr{L}(\Model(x,\hat{\theta}_{adaptive}), y)}}{\Expect{}{\mathscr{L}(\Model(x,\theta^*), y)}} - 1.
        \end{equation}

        $ALB$ is thus the proportion of expected risk in excess of what the best parameter value would deliver.

\begin{figure}[h!]
			\begin{subfigure}{.48\textwidth}
				\includegraphics[width=\linewidth]{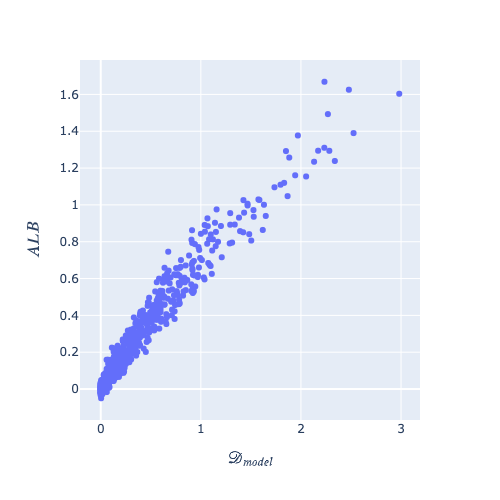}
				\caption{$\Model$ is linear, $\TrueConditional$ is quadratic.}
				\label{fig:linear-dmodel-12}
			\end{subfigure}\hfill\begin{subfigure}{.48\textwidth}
				\includegraphics[width=\linewidth]{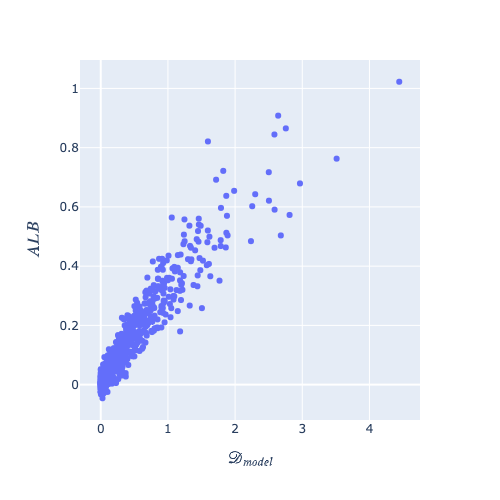}
				\caption{$\Model$ is quadratic, $\TrueConditional$ is cubic.}
				\label{fig:linear-dmodel-23}
			\end{subfigure}
			\caption[Linear regression: Empirical result 2.]{Linear regression: Empirical result 2 (The \emph{degree of} ALB depends on the \emph{degree of} misspecification).
        Points show each of 1,000 simulation experiments.
      }
			\label{fig:linear-dmodel}
		\end{figure}

        Figure \ref{fig:linear-dmodel} shows that $ALB$ varies positively with $\Misspecification$ (Equation \ref{eq:misspecification}).
        Figure \ref{fig:linear-dmodel-12} plots the $ALB$ values from the simulated experiments shown in Figure \ref{fig:linear-1} against the corresponding $\Misspecification$ values.\footnote{
            To calculate $ALB$, we approximated the risk by the $NLL$ of 100 observations from the target distribution.
            $\Expect{}{\mathscr{L}(\Model(x, \hat{\theta}_{adaptive}), y)}$ was calculated as the $NLL$ assigned by the estimated model after 100 steps of Bayesian adaptive design.
        }
        Figure \ref{fig:linear-dmodel-23} shows results from the same analysis performed on a series of simulated experiments designed to estimate a polynomial regression function misspecified as degree two when the generating function was degree three.\footnote{
            The distribution of generating parameters for the simulations shown in Figure \ref{fig:linear-dmodel-23} was chosen so the distribution of $\Misspecification$ values resembled the distribution of $\Misspecification$ values in Figure \ref{fig:linear-dmodel-12}.
        }
        In both cases, more misspecification predicts more ALB.

		As discussed in \S \ref{sec:measure-misspecification}, we expected the effect of $\Misspecification$ on $ALB$ would operate largely through its effect on the design distribution: The more mistaken the estimation process is about the generating model, the more mistaken it will be about where in the design space is most informative to query.
		However, note that the finding shown in Figure \ref{fig:linear-dmodel} \emph{cannot} stem from the effect of $\Misspecification{}$ on the distribution of designs.
		In \S \ref{sec:bayes-linreg}, we showed that the design distribution depends only the model structure and prior distributions --- which are held constant in the experiments shown within each panel of Figure \ref{fig:linear-dmodel}.
		Instead, this reflects that in this modeling paradigm, increased misspecification exacerbates risk stemming from sampling bias more generally --- i.e., irrespective of whether the bias was induced by active learning.\footnote{
		    We observed the same pattern for other restrictions on the design distribution, e.g., restrictions on the domain like that shown in Figure \ref{fig:misspecification}.
		}

		Intuition for this result may stem from Figure \ref{fig:misspecification}: With reference to the bottom right panel, a model class that allowed for non-linearity may not have been able to definitively rule out the best-fitting model on the basis of the observed data.
		Whether this occurred in practice would depend on, e.g., how this model class incorporated non-linearity.
		A complete explanation would need to take into account both the nature of the bias in the design distribution and the nature of the model misspecification --- an avenue for future work discussed in \S \ref{sec:future-work}.

    \subsection{Result 3: The presence of ALB depends on $\sigma$.}\label{sec:noise}
\begin{figure}
			\begin{subfigure}{.48\textwidth}
				\includegraphics[width=\linewidth]{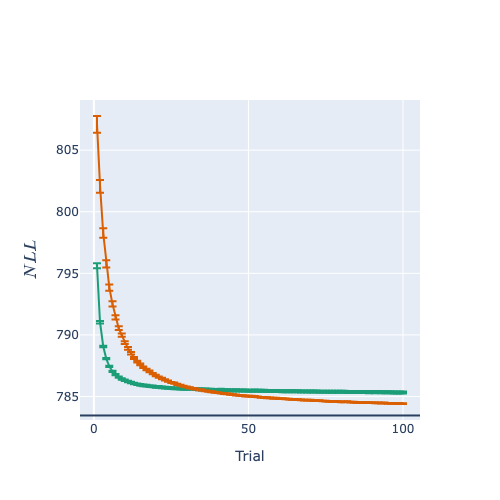}
				\caption{$\Model$ is linear, $\TrueConditional$ is quadratic.
				    Green lines ({\color{cb1} \rule{4mm}{1mm}}) show learning under the design distribution shown in Figure \ref{fig:designs-12-noise} (the Bayesian adaptive design).
				}
				\label{fig:linear-noise-12}
			\end{subfigure}\hfill\begin{subfigure}{.48\textwidth}
				\includegraphics[width=\linewidth]{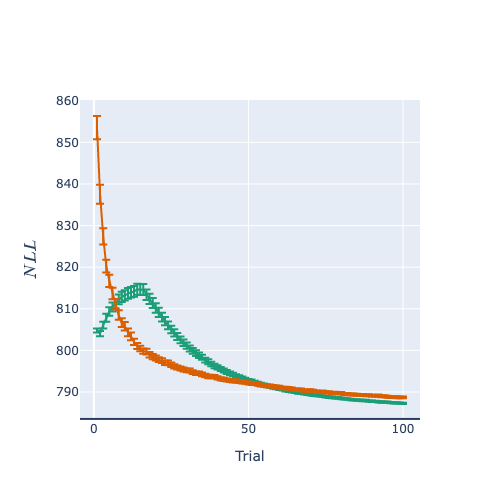}
				\caption{$\Model$ is quadratic, $\TrueConditional$ is cubic.
				    Green lines ({\color{cb1} \rule{4mm}{1mm}}) show learning under the design distribution shown in Figure \ref{fig:designs-23-noise} (the Bayesian adaptive design).
				}
				\label{fig:linear-noise-23}
			\end{subfigure}
			\begin{subfigure}{.48\textwidth}
				\includegraphics[width=\linewidth]{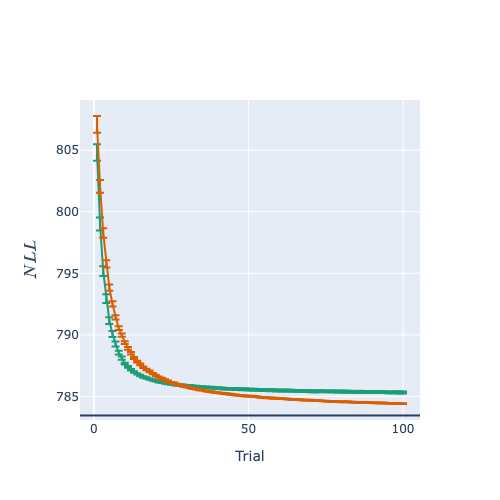}
				\caption{$\Model$ is linear, $\TrueConditional$ is quadratic.
				    Green lines ({\color{cb1} \rule{4mm}{1mm}}) show learning under the design distribution shown in Figure \ref{fig:designs-12}.
				}
				\label{fig:linear-noise-fixed-12}
			\end{subfigure}\hfill\begin{subfigure}{.48\textwidth}
				\includegraphics[width=\linewidth]{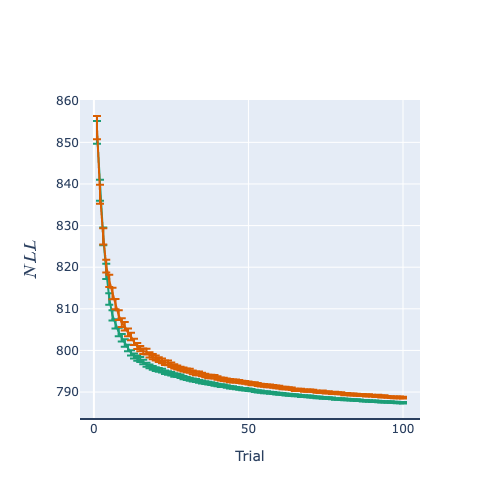}
				\caption{$\Model$ is quadratic, $\TrueConditional$ is cubic.
				    Green lines ({\color{cb1} \rule{4mm}{1mm}}) show learning under the design distribution shown in Figure \ref{fig:designs-23}.
				}
				\label{fig:linear-noise-fixed-23}
			\end{subfigure}
			\caption[Linear regression: Empirical result 3.]{Linear regression: Empirical result 3 (The presence of ALB depends on $\sigma$).
        Effect of increasing $\sigma$ on ALB.
			  As in Figure \ref{fig:linear}, but the additive noise term $\epsilon$ in the hypothesized model class has a standard deviation of 1,000 (as opposed to 100).
			  \textbf{(a,b)} Green lines ({\color{cb1} \rule{4mm}{1mm}}) show results using Bayesian adaptive design during model estimation, as before.
			  \textbf{(c,d)} As in \textbf{a} and \textbf{b}, but green lines ({\color{cb1} \rule{4mm}{1mm}}) correspond to learning from the distribution induced by active learning when $\sigma = 100$ (shown in Figures \ref{fig:designs-12} and \ref{fig:designs-23}).
			}
			\label{fig:linear-noise}
		\end{figure}

        Figures \ref{fig:linear-noise-12} and \ref{fig:linear-noise-23} show the evolution of risk for two misspecified model classes of the form shown in Figure \ref{fig:linear-dmodel}.
        However, instead of additive noise distributed $\mathscr{N}(0,100)$, the hypothesized model class incorporates additive noise distributed $\mathscr{N}(0, 1000)$.
        Figures \ref{fig:linear-noise-12} and \ref{fig:linear-noise-23} show that ALB is not present in these simulations: Adaptive sampling leads to an almost identical asymptotic risk as random sampling.
        In other words, the incorporation of this additional noise eliminates the ALB.\footnote{
            Note that $\theta^*$, which is used in the calculation of $ALB$, does not (here) depend on the degree of noise.
        }

	There are two reasons we might observe that higher $\sigma$ leads to lower $ALB$: 1) Higher $\sigma$ reduces $\Misspecification$, increasing robustness to sampling bias in general, and 2) higher $\sigma$ affects the distribution of designs, resulting in a \emph{design shift} that is more robust to misspecification.

	\paragraph{1. Higher $\sigma$ reduces $\Misspecification$.}
	    In the previous section, we reported our result that higher misspecification led to higher ALB.
	    In this context, increasing $\sigma$ systematically decreases the degree of model misspecification according to our measure: As the predictive distribution corresponding to the hypothesized model class becomes increasingly dispersed, it overlaps more with the data-generating distribution.
	    Lower divergence from $\TrueConditional$ to the hypothesized model class implies lower misspecification.
        Indeed, increasing $\sigma$ does lead to reduced model misspecification in our experiments: When $\sigma = 1,000$, the maximum $\Misspecification$ value we observed is 1.85 (compared to 4.44 when $\sigma = 100$).
        This in itself may be enough to eliminate ALB.

	\paragraph{2. Higher $\sigma$ shifts the design distribution.}
	    Because it is a direct intervention on the model structure, higher $\sigma$ \emph{does} affect the design distributions, unlike the sources of variation in $\Misspecification$ explored in the previous section.
	    Figure \ref{fig:designs} shows the $D$-optimal designs corresponding to models with $\sigma = 100$ vs. $\sigma = 1,000$.
	    While similar, the designs are not identical.
	    It is possible that the design distribution induced by $\sigma = 1,000$ is more robust to misspecification.

	To disentangle these two mechanisms, we ran a set of simulation experiments identical to those shown in Figures \ref{fig:linear-noise-12} and \ref{fig:linear-noise-23}, except that we fixed the optimal design distribution to the distribution induced by $\sigma = 100$.
	In effect, this eliminated the effect of the higher value of $\sigma$ on the design distribution, meaning that any reduction in $ALB$ could be attributed to $\sigma$'s effect on $\Misspecification$ (or other effects of $\sigma$ on robustness to sampling bias).

	The results of these experiments are shown in Figures \ref{fig:linear-noise-fixed-12} and \ref{fig:linear-noise-fixed-23}.
	Both panels show a reduction in $ALB$ nearly identical to that shown in Figures \ref{fig:linear-noise-12} and \ref{fig:linear-noise-23}, implying that the robustness to ALB is almost entirely due to the increased dispersion of the predictive distribution.

\section{Application to preference learning}\label{sec:additional-ex}
    The previous simulation results dealt with estimating linear regression models, which have convenient yet idiosyncratic properties (see \S \ref{sec:bayes-linreg} for discussion).
    In Appendix \ref{ap:classification}, we show that Results 1 and 3 generalize to a non-linear extension of this paradigm framed as a classification problem.\footnote{
        This artificial classification paradigm did not exhibit enough variation in $\Misspecification$ to test Result 2.
    }
    In this section, we show that all three results hold in another non-linear modeling paradigm with practical application.

    \emph{Preference learning} is a paradigm in which human subjects are presented with two options and asked to select one.
    Participants' stated preferences are used to recover a latent preference function that maps option attributes to how much they value that option, and so how likely they are to choose it.
    We hypothesized that misspecification of the preference function would lead to ALB.
    In addition, we hypothesized that increasing the amount of noise specified as part of the hypothesized model class --- in this case, how tightly values output by the preference function are linked to choices --- would reduce the ALB.

    In particular, we examine risky decision-making tasks, an area where Bayesian adaptive design has been used to select between competing psychological theories \citep{golovin_near-optimal_2010,cavagnaro_optimal_2013}.
    In this experimental paradigm, participants are presented with options between \textbf{gambles}, defined as discrete sets of outcomes with given probabilities and payoffs.
    For example, an input to a model of $\langle [.5, 10, .5, -10], 0 \rangle$ might indicate a choice between, on the one hand, a 50/50 chance of gaining or losing \$10, and on the other hand the surety of no change.

    We specified our design space as 200 randomly-generated gambles of the form $\langle [p, G, 1-p, L], 0 \rangle$ where the outcome $G$ was strictly greater than 0 and the outcome $L$ was strictly less than 0.
    The target distribution $\TargetDist$ was specified as a discrete uniform distribution across the design space.

\begin{figure}[h!]
      \begin{subfigure}{.48\textwidth}
	       \includegraphics[width=\linewidth]{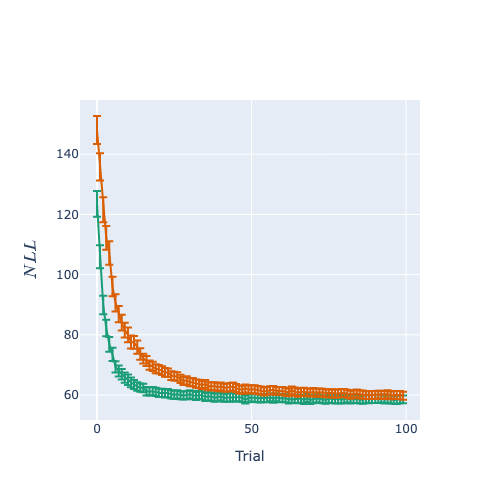}
	       \caption{\emph{Well-specified}: $\TrueConditional$ is an instance of EUT, and $\Model(x, \Theta)$ consists of instances of EUT.
	          The horizontal line shows the loss achieved by the true model, which is nearly achieved by both sampling procedures.
	      }
	      \label{fig:cpt-wellspecified}
      \end{subfigure}\hfill\begin{subfigure}{.48\textwidth}
	      \includegraphics[width=\linewidth]{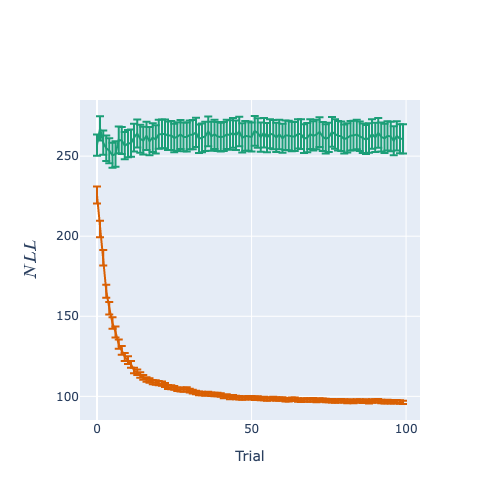}
	      \caption{\emph{Misspecified}: $\TrueConditional$ is an instance of CPT, yet $\Model(x, \Theta)$ contains instances of EUT.  {\\ ~ \\} }
	      \label{fig:cpt-misspecified}
      \end{subfigure}
      \begin{subfigure}{.48\textwidth}
	      \includegraphics[width=\linewidth]{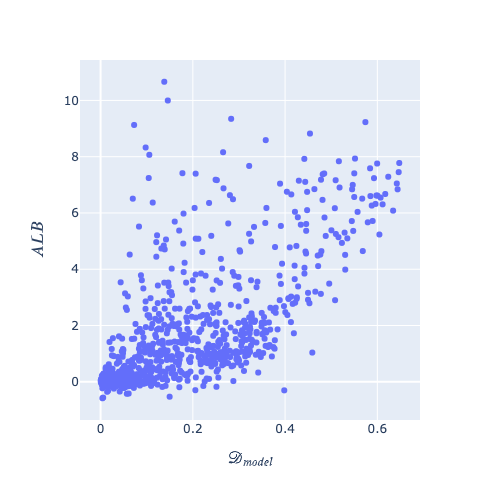}
	      \caption{$ALB$ as a function of $\Misspecification$ for the simulations in Figure \ref{fig:cpt-misspecified}.}
	      \label{fig:cpt-dmodel}
      \end{subfigure}\hfill\begin{subfigure}{.48\textwidth}
	      \includegraphics[width=\linewidth]{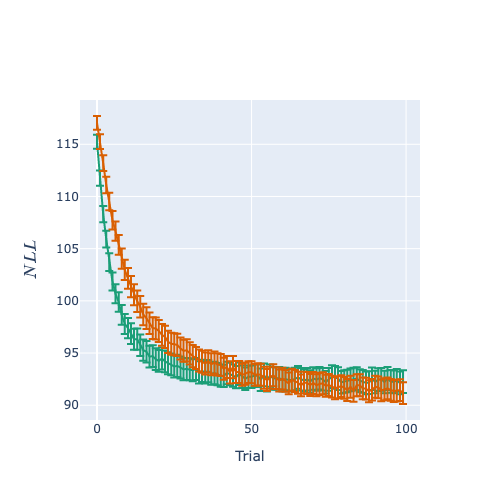}
	      \caption{\emph{Misspecified plus noise}: Same as Figure \ref{fig:cpt-wellspecified} but $\Model$ allows for more noise.}
	      \label{fig:cpt-noisy}
      \end{subfigure}
      \caption[Preference learning: Empirical results.]{Preference learning: Empirical results.
        Each panel shows risk incurred across 100 time steps of Bayesian updating.
        Risk is measured as the negative log likelihood ($NLL$) of responses to all 200 elements of the design space.
        Lines show the means across 1,000 simulations, and the error bars $\pm 1$ standard error about the mean.
        Green lines ({\color{cb1} \rule{4mm}{1mm}}) show results using Bayesian adaptive design during model estimation.
        Orange lines ({\color{cb2} \rule{4mm}{1mm}}) come from randomly sampling the target distribution.
      }
		\label{fig:cpt}
    \end{figure}

    We specified the hypothesized model class using expected utility theory (EUT), as in classical decision theory \citep{von_neumann_theory_1947}.
    Figure \ref{fig:cpt-wellspecified} compares adaptive and random sampling in estimating one instance of EUT: As anticipated, both methods eventually achieve the best-fitting $\theta$, with adaptive sampling converging faster, i.e., there is no ALB here.
    Details of the experimental set-up are provided in Appendix \ref{ap:cpt}.

    Figure \ref{fig:cpt-misspecified} compares adaptive and random sampling when the data-generating process follows cumulative prospect theory (CPT), a better descriptor of how individuals make similar decisions than EUT \citep{tversky_advances_1992}.
    In each experiment, the parameters of CPT are set so that they cannot be mimicked by any instance of EUT, guaranteeing misspecification.
    As expected, we observe ALB: adaptive sampling leads to systematically higher risk than random sampling.
    Figure \ref{fig:cpt-dmodel} shows that as in the linear regression paradigm, higher $\Misspecification$ generally predicts higher $ALB$ (although the relationship isn't as strong).\footnote{
        Since $\theta^*$ can't be analytically solved for in this example, we approximated $\Misspecification$ using the $D_{KL}$ from the true distribution to the predictive distribution $\Model(x, \hat{\theta})$.
    }

    Figure \ref{fig:cpt-noisy} has the same setup as Figure \ref{fig:cpt-misspecified}, but the hypothesized model assumes choices are noisier than they truly are (i.e., the function linking the output of the latent preference function to choices induces more stochasticity).
    As in Figure \ref{fig:linear-noise}, introducing noise eliminates the ALB.\footnote{
        In many instances of this paradigm, the hypothesized model class has a parameter corresponding to the ``noisiness'' of observed choices \citep{broomell_parameter_2014}.
        While in our example the model specification fixes this parameter, future work should look at what happens when the noise parameter has a non-degenerate prior distribution.
    }

\section{Discussion}\label{sec:discussion}
    Our work investigated when active learning bias afflicts Bayesian adaptive experimental design.
    We primarily drew insight from results in a class of polynomial regression models.
    While the tractability and interpretability of these models made them an appealing paradigm in which to begin our analysis, they share particular idiosyncracies that may affect the generalizability of these insights (e.g., the equivalence between the batch and adaptive experimental design solutions).
    Our second example investigated the behavior of non-linear models, extending the insights from the linear case to a class of models used for online recovery of user preferences.
    In both paradigms, we (1) demonstrated that the presence of active learning bias depends on model misspecification, (2) developed a measure of misspecification for probabilistic model classes, (3) showed that the extent of active learning bias varies with the degree of misspecification, and (4) showed that the amount of observational noise specified by the model class mitigates the extent of active learning bias.

    As demonstrated by our second set of empirical results, such biases can lead to severely degraded prediction performance in the context of active learning of user preferences.
    Beyond their effects on prediction performance, biased parameter estimates can themselves have unintended consequences.
    As discussed in \S \ref{sec:al}, a common use case for Bayesian adaptive design is the design of experiments for individualized assessment and scientific inference.
    In such cases, parameters often correspond to scientifically-meaningful and interpretable quantities, such as propensity for risk-seeking behavior \citep{cavagnaro_optimal_2013} or impulsivity \citep{ahn_rapid_2020}.
    The presence of biases in these inferences can result in misinformed institutional and individual decisions, such as suboptimal treatment recommendations.

	\subsection{Recommendations}
	    Users of Bayesian adaptive design who wish to avoid erroneous inferences and predictions should work to ensure that their model classes are well-specified, such as by applying existing tests for misspecification \citep{white_maximum_1982}.
        Failing that, our results suggest ways users could reduce the anticipated active learning bias, such as increasing the amount of noise --- the amount of inherent variance in outputs --- specified by the model class.
        Of course, this robustness comes at the cost of weakening the inferences one can draw from finite data.
        We recognize that researchers must perform a delicate balancing act between specifying adequate uncertainty, and potentially forgoing the cost savings and statistical power resulting from adaptive experimentation in cases where existing models are not too misspecified.
        Existing techniques that adaptively reduce the degree of inherent variance, such as active learning with Gaussian processes (GPAL) \citep{houlsby_collaborative_2011,chang_data-driven_2021}, are promising methods for achieving this balance.

        However, these more flexible techniques often forgo the interpretability of parameter estimates usually associated with more constrained modeling paradigms.
        Experimenters could aim to achieve the ``best of both worlds'' by setting aside resources to run multiple experiments that combine flexible modeling paradigms like GPAL with visual inspection and model respecification in an iterative model development process \citep{chang_data-driven_2021}.
        In addition or instead, experimenters can adjust the sampling strategy itself in a way that leverages the power of adaptive designs \emph{and} protects against biased design distributions, by, e.g., incorporating more ``exploratory'' trials \citep{box_basis_1959,osugi_balancing_2005,ali_active_2014}.

    \subsection{Directions for future work}\label{sec:future-work}
        Our conclusions in this paper hinge on the results of simulation experiments.
        In the future, theoretical results should delineate the conditions under which one should expect active learning bias, as, e.g., a function of $\Misspecification$.
        Such a study could perhaps draw from theoretical results on the effect of misspecification on the effectiveness of parameterized bandit learning algorithms, a conceptually similar class of active learning methods \citep{kannan_smoothed_2018,bogunovic_misspecified_2021}.

        We hope that such continued work will generate a more complete understanding of how and why decreasing $\Misspecification$ and increasing the degree of noise enhances robustness to ALB in the two modeling paradigms presented in this paper --- and the extent to which this effect can be generalized to other modeling paradigms and active learning rules.
        As discussed in \S \ref{sec:measure-misspecification}, we expected this robustness to stem from the effect of $\Misspecification$ on the design distribution.
        However, as demonstrated in \S \ref{sec:degree-misspecification} and \S \ref{sec:noise}, higher $\Misspecification$ increases learned bias independent of the source of bias in the design distribution.
        Understanding these results in the context of the relationship between $\Misspecification$ and robustness to covariate shift more generally is a promising avenue for future research.

\section*{Acknowledgments}
    This work benefited enormously from discussion with Marina Dubova and Daniel Cavagnaro, and comments from several anonymous reviewers.
    SJS was supported by a Tata Consultancy Services (TCS) Fellowship while contributing to this work.

\bibliographystyle{apacite}
\bibliography{preprint_v2}

\appendix

\section{Derivation of Equation \ref{eq:linear-Eu}}\label{ap:linear-Eu}
    As discussed in \S \ref{sec:baedo}, $\Utility{}$ is equivalent to the mutual information between $\cond{y}{x}$ and $\beta$, which can equivalently be written as:

    \begin{align}
        \Utility{} &= I(\cond{y}{x} ; \beta) \nonumber \\
        &= H_t(\cond{y}{x}) - H_t(\cond{y}{x}, \beta) \nonumber
    \end{align}

    where $H_t$ is the entropy function taken with respect to the indicated posterior distribution at time $t$, and $H_t(\cond{y}{x}, \beta)$ is the conditional entropy of $\cond{y}{x}$ given $\beta$.

    $y \sim \mathscr{N}(M_t\phi^k(x)^T, \sigma^2 + \phi^k(x)^TS_t\phi^k(x))$, so $H_t(\cond{y}{x})$ can be computed using the equation for the entropy of a normal distribution, as follows:

    \begin{align}\label{eq:linear-hy}
        H_t(\cond{y}{x}) &= \frac{1}{2}\log{\left( 2\pi(\sigma^2 + \phi^k(x)^TS_t\phi^k(x)) \right)} + \frac{1}{2} \nonumber
    \end{align}

    $H_t(\cond{y}{x}, \beta)$ is computed as follows:

    \begin{align}
        H_t(\cond{y}{x}, \beta) &= - \int_\beta{ \int_y{ \log{\left( \Model(\cond{y}{x}, \beta) \right)} ~ \Model(\cond{y}{x}, \beta) ~ \Posterior_t(\beta)  ~ dy} ~ d\beta} \nonumber \\
        &= \int_\beta \int_y \log{\left( \sigma\sqrt{2\pi} \right)} + \frac{1}{2} \left( \frac{y - \beta \phi^k(x)^T}{\sigma} \right)^2 ~ \Model(\cond{y}{x}, \beta) ~ \Posterior_t(\beta) ~ dy ~ d\beta \nonumber \\
        &= \log{\left( \sigma\sqrt{2\pi} \right)} + \frac{1}{2\sigma^2} \int_\beta \int_y (y - \beta \phi^k(x)^T)^2 ~ \Model(\cond{y}{x}, \beta) ~ \Posterior_t(\beta) ~ dy ~ d\beta \nonumber \\
        &= \log{\left( \sigma\sqrt{2\pi} \right)} + \frac{1}{2\sigma^2} \int_\beta \sigma^2 ~ \Posterior_t(\beta) ~ d\beta \nonumber \\
        &= \log{\left( \sigma\sqrt{2\pi} \right)} + \frac{1}{2} \nonumber
    \end{align}

    We can then write the complete expression for $\Utility{}$:

    \begin{align}
        \Utility{} &= \left( \frac{1}{2}\log{\left( 2\pi(\sigma^2 + \phi^k(x)^TS_t\phi^k(x)) \right)} + \frac{1}{2} \right) - \left( \log{\left( \sigma\sqrt{2\pi} \right)} + \frac{1}{2} \right) \nonumber \\
        &= \frac{1}{2}\log{\left( \sigma^2 + \phi^k(x)^TS_t\phi^k(x) \right)} - \log{\left( \sigma \right)} \nonumber
    \end{align}

    as stated in Equation \ref{eq:linear-Eu}.

\section{Replication of Results 1 and 3 in an artificial classification problem}\label{ap:classification}
    This problem retains the underlying structure of the regression example presented in \S \ref{sec:linear-ex}.
    However, as in the preference learning example presented in \S \ref{sec:additional-ex}, the dependent variables follow a Bernoulli distribution.
    While this problem provides a conceptual ``bridge'' between the linear regression and preference learning paradigms, we defer these results to the appendix because they are based on fewer experiments, and because the modeling paradigm itself is of less theoretical and practical interest.\footnote{
        Because the parameter space is both multi-dimensional and does not admit a closed-form representation, the computational cost of each experiment was higher than in the experiments presented in \S \ref{sec:linear-ex} and \S \ref{sec:additional-ex}.
        In addition, only a small fraction of parameter settings we investigated induced misspecification.
        For these reasons, we ran fewer simulation experiments under this modeling paradigm.
    }

    For this problem, the hypothesized model class consisted of degree-one polynomials linked to binary observations by a logistic link function --- i.e., posited that $y|x \sim \mathrm{Bernoulli}(\frac{1}{1 + e^{-\epsilon(\beta_0 + \beta_1 x)}})$.
    $\epsilon$ determines the degree of observational noise in the hypothesized model class, and was fixed within each model class.
    Observations were generated by the same degree-two polynomial regression model described in \S \ref{sec:linear-ex}, also pushed through a logistic link function, i.e., observations were distributed $y|x \sim \mathrm{Bernoulli}(\frac{1}{1 + e^{-(\beta_0 + \beta_1 x + \beta_2 x^2)}})$.
    Note that in the true model, the degree of observational noise does \emph{not} vary.
    This asymmetry between the specification of observational noise in the hypothesized model classes and true models will allow us to test Result 3.\footnote{
        Parameter settings for the generating model were selected from among the parameter settings used for the corresponding regression problem to maximize $\Misspecification$ when $\epsilon = 1$.
    }

\begin{figure}[h!]
        \begin{subfigure}{.32\linewidth}
	        \includegraphics[width=\linewidth]{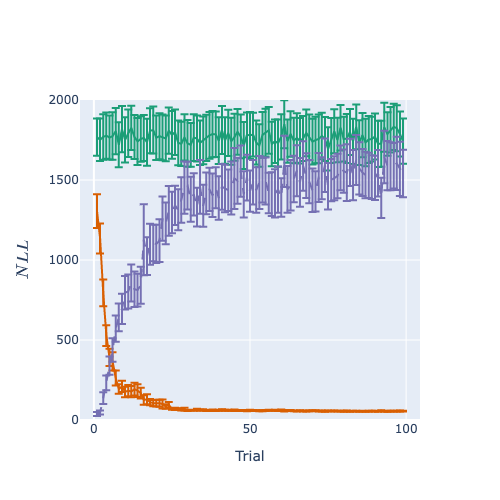}
	        \caption{$\epsilon = 1$.}
	    \end{subfigure}\hfill\begin{subfigure}{.32\linewidth}
	        \includegraphics[width=\linewidth]{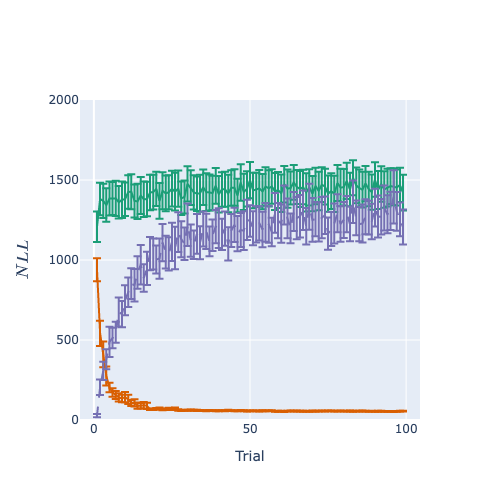}
	        \caption{$\epsilon = .1$.}
	    \end{subfigure}\hfill\begin{subfigure}{.32\linewidth}
	        \includegraphics[width=\linewidth]{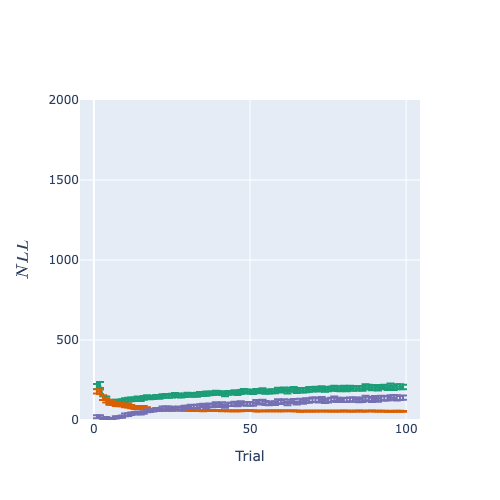}
	        \caption{$\epsilon = .01$.}
	    \end{subfigure}
	    \caption[Classification: Empirical results.]{Classification: Empirical results.
        Each panel shows risk over 100 time steps of Bayesian updating for the artificial classification model.
		    The underlying $\TrueConditional$ is a quadratic logistic regression.
		    The hypothesized model class consists of linear logistic regressions, where the linear regression output is multiplied by a factor $\epsilon$ before being pushed through the link function.
		    Lines are means across 1,000 simulated experiments, with error bars showing $\pm 1$ standard error around the mean.
		    Green lines ({\color{cb1} \rule{4mm}{1mm}}) show results using Bayesian adaptive design optimization during model estimation.
		    Orange lines ({\color{cb2} \rule{4mm}{1mm}}) show results using random sampling from the target distribution.
		    Purple lines ({\color{cb3} \rule{4mm}{1mm}}) show $50 \times \left( \frac{\Expect{}{\mathscr{L}(m, \hat{\theta}_{adaptive})}}{\Expect{}{\mathscr{L}(m, \hat{\theta})}} - 1 \right)$ where the risk is calculated at the indicated time step.
	    }
		\label{fig:binary}
	\end{figure}

    Figure \ref{fig:binary} shows that classification follows the same trend as regression: In all three panels, each of which corresponds to a different value of $\epsilon$, adaptive sampling leads to higher risk than random sampling.\footnote{
        Unlike the previous examples, the posterior parameter distributions in this example do not have a closed-form analytical solution.
        We approximated expectations under the parameter distribution at each time step $t$ using importance sampling, where the importance distribution was a weighted kernel density estimate from 10,000 samples from the parameter distribution at $t-1$, each sample weighted by the product of its importance weight at time $t-1$ and its likelihood under the observation made at $t$.
    }

	Figure \ref{fig:binary} also shows that Result 3 holds in this modeling paradigm: Higher degrees of hypothesized observational noise (lower $\epsilon$) leads to lower ALB.\footnote{
	    We found that the $\Misspecification$ values in this problem did not exhibit adequate variation to assess the applicability of Result 2 (\S \ref{sec:degree-misspecification}; higher $\Misspecification$ predicts higher ALB).
	}
	The purple lines show the degree of ALB: They are equal to $50 \times \left( \frac{\Expect{}{\mathscr{L}(m, \hat{\theta}_{adaptive})}}{\Expect{}{\mathscr{L}(m, \hat{\theta})}} - 1 \right)$ with the risk calculated at the indicated time step.
	At $t = 100$, this is a good approximation of $50 \times ALB$ (with the factor of 50 included to put these values on roughly the same scale as the absolute risk).
	Figure \ref{fig:binary} shows that as $\epsilon$ decreases, i.e., the degree of hypothesized noise increases, the degree of ALB also decreases (although is in no case eliminated, as in the experimental results presented in \S \ref{sec:linear-ex} and \S \ref{sec:additional-ex}).

\section{Details of preference learning example}\label{ap:cpt}
    For each input of the form $\langle [p, G, 1-p, L], 0 \rangle$, we refer to the option that yields $G$ with probability $p$ and $L$ with probability $1-p$ as the \emph{gamble}, and model $P_{gamble}$, the probability that a participant selects the gamble over the certain outcome of no payoff.

    We specified both EUT and CPT such that $P_{gamble}$ is a monotonic function of $V(gamble)$, the participant's valuation of, or strength of preference for, the gamble over the certain outcome.\footnote{
        Classical expected utility theory prescribes deterministic choices \citep{von_neumann_theory_1947}, but behavioral scientists have found that people's probability of selecting a given option tends to increase with their strength of preference for it; see \citet{stott_cumulative_2006} for a review.
        In general, we use ``EUT'' and ``CPT'' to refer to the mathematical specifications we used in our simulations, and note where these do not exactly align with the claims of the theoretical framework of Expected Utility Theory or Cumulative Prospect Theory, respectively.
    }
    We used the following \emph{choice function} to link $V(gamble)$ to $P_{gamble}$ \citep{stott_cumulative_2006}:

    \begin{equation}
        P_{gamble} = \frac{1}{1 + e^{-\epsilon V(gamble)}}.
    \end{equation}

    The function mapping $[p, G, 1-p, L]$ to $V(gamble)$ depends on whether the model class is EUT or CPT.
    $\epsilon$ is a ``sensitivity'' parameter that determines how deterministically $V(gamble)$ is linked to a participant's observed preferences.
    When $\epsilon$ is large, participants are more likely to select options to which they assign a higher valuation.
    We introduce ``noise'' into the hypothesized model class by manipulating $\epsilon$ (see \S \ref{ap:cpt-params}).

    \subsection{Specification of EUT}
        While the theoretical framework of EUT as a decision theory is consistent with a variety of functional forms \citep{von_neumann_theory_1947}, for the sake of this example, we use a deliberately simplified version where $V(gamble)$ is determined by a single parameter $\alpha$:

        \begin{equation}
            V(gamble) = p G^\alpha - (1-p) (-L)^\alpha.
        \end{equation}

        When estimating $\alpha$, we assigned it a standard uniform prior, and represented its distribution as a grid of 1,001 points.

    \subsection{Specification of CPT}
        We used a simplified version of CPT, which is a special case of the full theory in case an option contains only one positive and one negative outcome.
        Interested readers can consult \citet{tversky_advances_1992} or \citet{broomell_parameter_2014} for discussion of the general formulations of CPT.

        CPT differs from EUT in two important ways \citep{tversky_advances_1992}:
        \begin{enumerate}
            \item In EUT, the \emph{decision weight} associated with a payoff, or its coefficient in the linear combination of payoffs used to calculate $V(gamble)$, is simply the probability of its realization.
            In CPT, the decision weight is a non-linear function of the payoff's associated probability, $w(\cdot)$.
            \item In CPT, payoffs below 0 are additionally multiplied by a \emph{loss aversion} term $\Lambda$.
        \end{enumerate}

        In other words, CPT posits that $V(gamble)$ is computed as:

        \begin{equation}
            V(gamble) = w(p)G^{\alpha} - w(1-p)\Lambda(-L)^{\alpha}.
        \end{equation}

        We used a probability weighting function $w$ proposed by \citet{prelec_probability_1998}:

        \begin{equation}
            w(p) = e^{-(-ln(p))^\gamma}.
        \end{equation}

        CPT thus has three free parameters: $\alpha$, $\gamma$ and $\Lambda$.

    \subsection{Parameter settings}\label{ap:cpt-params}
        In Figure \ref{fig:cpt-wellspecified}, $f$ is EUT, with the generating value of $\alpha$ in each experiment drawn from a standard uniform distribution.
        In Figures \ref{fig:cpt-misspecified} and \ref{fig:cpt-noisy}, $f$ is CPT, with the generating values of $\alpha$ and $\gamma$ in each experiment each drawn from a standard uniform distribution, and the generating value of $\Lambda$ drawn from a $\mathrm{Unif}(1, 2)$ distribution.

        In Figures \ref{fig:cpt-wellspecified} and \ref{fig:cpt-misspecified}, $\epsilon = 1$ for both the true and hypothesized model class.
        In Figure \ref{fig:cpt-noisy}, $\epsilon = 1$ for the true model but $\epsilon = .1$ for the hypothesized model class.

\end{document}